# Price-Volume Correlations Are Always Zero

Victor Olkhov
Independent, Moscow, Russia
victor.olkhov@gmail.com
ORCID: 0000-0003-0944-5113

## Abstract

We argue that the use of the empirical time series of values, volumes, and prices of consecutive trades made during the averaging interval to calculate the price-volume correlation in a way that follows the definition of correlation always gives zero correlation. We show that the usual expression for the empirical calculation of the correlation misleads the researchers and doesn't fulfill the definition of price-volume correlation. The mistake has its origin in the dependence of volume weighted average price on the mean values and volumes of trades during the averaging interval but not on time series of prices only. The true sense of the declared positive or negative results of the empirical calculations of price-volume correlations made by numerous researchers in reality is determined by the difference between the volume weighted average price and its limiting expression of average price, which is correct only when all volumes of consecutive trades are constant, and multiplication of this difference by the mean trade volume. To empirically determine the statistical dependence between time series of prices and trade volumes, one should calculate the correlations between prices and squares of trade volumes or correlations between squares of prices and squares of volumes. We derive the expressions for empirical calculations of these correlations.

---

This research received no support, specific grant, or financial assistance from funding agencies in the public, commercial, or nonprofit sectors.

## 1. Introduction

The empirical evidence of positive or negative price-volume correlations was presented by numerous researchers (Tauchen and Pitts, 1983; Karpoff, 1987; Gallant et al., 1992; Stein, 1993; DeFusco et al., 2017; DeFusco et al., 2022) and others.

Meanwhile, the correlation is not measured but calculated only. To calculate the price-volume correlation, one should use the empirically measured time series of the values, volumes, and prices of trades made at the exchange or other markets during a particular period and apply simple but important economic assumptions and models of probability theory. One should account for that different economic assumptions and approximations of the real market trades made during the averaging period may result in different methods and expressions for the empirical calculations of price-volume correlations.

In this paper we reconsider the "empirical evidence" of positive or negative price-volume correlations and show that in reality that "empirical evidence" describes statistical properties that are far different from correlations. We give simple proofs that the economically justified price-volume correlations are always zero. Zero price-volume correlations don't mean the lack of statistical dependence between price-volume time series. To find out the statistical dependence of these time series, one should calculate the empirical correlations between prices and squares of trade volumes or correlations between squares of prices and squares of volumes. We derive the expressions for these correlations.

## 2. Market-based origin of price-volume time series

We consider time series of values, volumes, and prices of consecutive trades made at the exchange or other markets during the averaging interval $\Delta$ (A.1) that can be equal to hours, days, etc. One may consider high-frequency trades made at the exchange with a tick period between consecutive trades. This is the highest frequency of trades in the economy. However, such a high frequency is not necessary and not too useful for the assessments of statistical moments of prices and volumes and their correlation during the averaging intervals equal to hours, days, etc. We don't require the use of high-frequency trade time series. We propose to consider the sums of the values and volumes of trades made at the exchange or other markets during each consecutive period $\varepsilon$ (A.1) that can be equal to minutes, hours, days, etc. as the values $C(t_i)$ and volumes $U(t_i)$ of consecutive trades made at time $t_i$ during the averaging interval $\Delta$ (A.1). We assume that the relation $\Delta = N \cdot \varepsilon$ holds and propose that the number $N$ of consecutive trades during $\Delta$ is rather high to obtain reliable estimates of statistical moments of trade values, volumes, and price-volume correlations in particular. For example, if *N~50-100*, then to assess price-volume correlations during the averaging interval *Δ~1 day (8 hours),* one should consider the period *ε~5-10 min.*

## 3. Market-based origin of VWAP

We focus our study on the market-based origin and the random changes of the values $C(t_i)$ and volumes $U(t_i)$ of consecutive trades at time $t_i$ during the averaging interval $\Delta$. The relations between the value $C(t_i)$, volume $U(t_i)$, and price $p(t_i)$ of a trade made at time $t_i$ during the averaging interval $\Delta$ are determined by a trivial trade equation (A.2). The randomness of trade values and volumes determines the stochastic properties of trade price. Actually, the equation (A.2) determines the dependence of statistical moments of price, and price-volume correlation in particular, on statistical moments of values and volumes. For convenience we denote mean values $C(t)$ and mean volumes $U(t)$ (A.5) of trades during $\Delta$ (A.1). As $C_\Sigma(t)$ and $U_\Sigma(t)$ (A.4), we denote the total value and volume of trades made during the averaging interval $\Delta$. In financial economics it is conventional to refer to Berkowitz et al. (1988) as the authors of volume weighted average price (VWAP) $p(t)$ (3.1; A.6).

$$p(t) = E[p(t_i)] = \frac{C_\Sigma(t)}{U_\Sigma(t)} = \frac{C(t)}{U(t)} = \frac{1}{\sum_{i=1}^{N} U(t_i)} \sum_{i=1}^{N} p(t_i) \cdot U(t_i) \tag{3.1}$$

$E[..]$ denotes mathematical expectation. Actually, the averaging of trade prices $p(t_i)$ while trade volumes $U(t_i)$ are irregular or random during the averaging interval was studied a long before Berkowitz et al. (1988). We refer to Diewert (1988, 1995), who gave a great review of methods for averaging prices and price indices that were developed almost for two centuries. We highlight the definition of average price: "..the prices should be weighted according to the relative mass-quantities that were sold at them." (Walsh, 1901, p. 96; Diewert, 1995, p. 21) and "..the average price found by dividing the total sum (or the full value) by the total quantities." (Walsh, 1921, p. 88). That is exactly the definition of VWAP (Berkowitz et al., 1988). Moreover, Walsh (1901, p. 97) outlines that the expression of the average price $p(t)$ (3.1) was proposed by Segnitz (1870) and refers to Heinz (1876), who claimed that ".. this method is employed in the statistical bureaus of several German states – Würtemberg, Bavaria, Gotha, and partly in Prussia and Hannover." (Walsh, 1901, p. 97). Hence, one should accept that the authorship of the definition of VWAP should be moved from Berkowitz et al. (1988), back for more than a century to Segnitz (1870), the researchers of "statistical bureaus of several German states," and Walsh (1901; 1921).

The definition of VWAP *p(t)* (3.1) accounts for the impact of the irregular or random volumes $U(t_i)$ of consecutive trades made during the averaging interval *Δ* (A.1). Moreover, VWAP *p(t)* (3.1) determines an important but limited case of trades made during *Δ* when all volumes of the successive trades are constant, $U(t_i)=U,\ i=1,..N$. If so, *p(t)* (3.1) takes the form *π(t)* (3.2; A.7)

$$if\, U(t_i) = U \ \ then \ \ p(t) = \frac{1}{\sum_{i=1}^{N} U(t_i)} \sum_{i=1}^{N} p(t_i) \cdot U(t_i) = \frac{1}{N \cdot U} \sum_{i=1}^{N} p(t_i) \cdot U = \frac{1}{N} \sum_{i=1}^{N} p(t_i) = \pi(t) \quad (3.2)$$

We highlight that the expression of the average price *π(t)* (3.2) has the economic meaning only in the limited case when all volumes of the consecutive market trades made during the averaging interval are constant.

## 4. Price-volume correlations of successive market trades

The definition of correlation between the prices $p(t_i)$ and volumes $U(t_i)$ is simple (4.1; A.8):

$$corr[p, U] = E[(p(t_i) - E[p(t_i)]) \cdot (U(t_i) - E[U(t_i)])] = E[p(t_i) \cdot U(t_i)] - E[p(t_i)]E[U(t_i)] \quad (4.1)$$

One may derive economically justified correlation (4.1) for two cases. The simplest case describes constant volumes $U(t_i)=U$ of the consecutive trades during the averaging interval. For this case, it is obvious that the correlation (4.1) $corr[p,U]=0$.

In the general case, the volumes $U(t_i)$ of consecutive trades are irregular or random during the averaging interval *Δ* (A.1). Then, the mean price $E[p(t_i)]$ in (4.1) should take the form of VWAP *p(t)* (3.1) and the use of the trade equation (A.2) and (A.4; A.5), give:

$$E[p(t_i) \cdot U(t_i)] = E[C(t_i)] = E[p(t_i)]E[U(t_i)] \quad (4.2)$$

From (4.1; 4.2), obtain that in the general case the price-volume correlation is always zero:

$$corr[p, U] = E[p(t_i) \cdot U(t_i)] - E[p(t_i)]E[U(t_i)] \equiv 0 \quad (4.3)$$

***4.1 The origin of the fallacy of the "empirical" positive or negative price-volume correlation***

However, numerous researchers provided the "empirical evidence" of positive or negative price-volume correlations. How were such assessments derived? We should reiterate that correlation is not measured empirically. Price-volume correlation can be calculated using the empirical time series of values $C(t_i)$, volumes $U(t_i)$, and prices $p(t_i)$ of consecutive trades measured during the averaging interval. To calculate the correlation between empirical time series of prices and volumes, one should use the price-volume time series in a manner that matches the definition of correlation (4.1), especially the right side of the definition of correlation (4.1). We believe that at this point "empirical researchers" made a substantial mistake that mislead their assessments of price-volume correlation. The origin of the flaw is following. The "empirical researchers" use the conventional expression of correlation $corr[x,y]_e$ (4.4; B.2) between two time series of random variables $x_i$ and $y_i$ that are presented in almost all manuals on statistical analysis of time series (Tsay, 2005, pp. 17; 25; Montgomery et al., 2015, pp. 38, 437):

$$corr[x,y]_e = \frac{1}{N}\sum_{i=1}^{N}(x_i - E[x]) \cdot (y_i - E[y]) \tag{4.4}$$

The above relations are so common and verified by so many studies that the "empirical researchers" apply them to the assessment of price-volume correlation without any caution. However, as we show in App. B. (B.2-B.10), the use of the well-known expressions (4.4; B.6) for the calculations of correlation between time series of random variables $x$ and $y$ is correct if and only if the mean values $E[x]$ and $E[y]$ obey (B.4). Only in this case do the relations (4.4) conform with the definition of correlation given in the right side of (4.1). While the mean volume $U(t)$ (A.5) obeys (B.4), the VWAP $p(t)$ (3.1; A.6) does not. This leads to the fact that the well-known expression (4.4) doesn't correctly define price-volume correlation as it is determined on the right side of (4.1). Indeed, the right side of (4.1) defines the price-volume correlation $corr[p,U]$ (4.5):

$$corr[p,U] = E[p(t_i) \cdot U(t_i)] - p(t)U(t) \tag{4.5}$$

Due to the equation (A.2) and the definition of mean value $C(t)$ (A.5), (4.5) takes the form:

$$corr[p,U] = \frac{1}{N}\sum_{i=1}^{N} p(t_i)U(t_i) - p(t)U(t) \tag{4.6}$$

However, the calculations of $corr[p,U]_e$ via (4.4) result in a different relation (4.7):

$$corr[p,U]_e = \frac{1}{N}\sum_{i=1}^{N}(p(t_i) - p(t)) \cdot (U(t_i) - U(t)) = \frac{1}{N}\sum_{i=1}^{N} p(t_i)U(t_i) - \pi(t)U(t) \tag{4.7}$$

The factor $\pi(t)$ (3.2; A.7) in the right side of (4.7) causes (4.7) to violate the definition of correlation (4.5; 4.6). The expression of $corr[p,U]_e$ (4.7) that was calculated via (4.4) doesn't determine the correlation between prices and volumes but, due to (3.1; B.9), results in (4.8; B.10):

$$corr[p,U]_e = [p(t) - \pi(t)]U(t) \tag{4.8}$$

Thus, the "empirical evidence" of positive or negative price-volume correlation" $corr[p,U]_e$ (4.7; 4.8) declared by many researchers in fact describes the difference between the VWAP $p(t)$ (3.1) and its limiting case $\pi(t)$ (3.2; A.7), which has the economic meaning of mean price only if all volumes of consecutive trades are constant.

It is evident that zero price-volume correlations don't mean the lack of statistical dependence between time series of prices and volumes. However, to assess it, one should empirically calculate correlations (see App. C) between prices and squares of trade volumes $corr[p,U^2]$ (C.8), or correlations between squares of prices and squares of trade volumes $corr[p^2,U^2]$ (C.16). As we show in App. C., these correlations obey simple relations (C.16):

$$corr[p^2, U^2] = 2corr[p, U^2] \cdot p(t) \tag{4.9}$$

To calculate the correlations $corr[p,U^2]$ (C.8) or $corr[p^2,U^2]$ (4.9; C.16) using empirical time series of the values $C(t_i)$, volumes $U(t_i)$, and prices $p(t_i)$ of the consecutive trades during $\Delta$ (A.1), one should follow the expressions (A.5; A.6; C.1; C.5; C.6).

## 5. Conclusion

The definition of correlation (4.1) between prices and trade volumes (4.2; 4.5; 4.6), VWAP $p(t)$ (3.1; A.6), the equation (A.2), and the definition of mean value $C(t)$ and volume $U(t)$ (A.5) result in zero price-volume correlation (4.3). The price-volume correlation for the limiting case of constant trade volumes and average price $\pi(t)$ (3.2; A.7) is all the more equal to zero.

The dependence of VWAP price $p(t)$ (3.1; A.6) on mean value $C(t)$ and volume $U(t)$ (A.5) causes that the well-known common expression (4.4) for the calculation of the correlation between empirical time series of prices and volumes doesn't match the definition (4.1) of price-volume correlation (4.5; 4.6) and actually describes the difference (4.8) between the VWAP $p(t)$ (3.1; A.6) and its limiting case $\pi(t)$ (3.2) that has the meaning of average price only when all volumes of consecutive trades are constant. The difference (4.8) between VWAP $p(t)$ (3.1; A.6) and $\pi(t)$ (3.2; A.7) is the origin of the declared positive or negative "empirical correlations" between prices and trade volumes.

To empirically assess the statistical interdependence between time series of prices and trade volumes, the researcher may calculate the correlations $corr[p,U^2]$ (C.8) or $corr[p^2,U^2]$ (C.16; 4.9).

## Appendix A. Price-Volume Correlation

We consider time series of values $C(t_i)$ and volumes $U(t_i)$ of consecutive trades made at time $t_i$, $i=1,..N$, during the averaging interval $\Delta$ (A.1):

$$\Delta = \left[t - \frac{\Delta}{2}; t + \frac{\Delta}{2}\right] \quad ; \quad t_i \in \Delta \quad ; \quad t_{i+1} = t_i + \varepsilon \quad ; \quad \Delta = N \cdot \varepsilon \tag{A.1}$$

As the values $C(t_i)$ and volumes $U(t_i)$, we consider the total values and volumes of all trades made at the exchange or other markets during the period $(t_{i-1}, t_i)$, $\varepsilon = t_i - t_{i-1}$. The period $\varepsilon$ can be equal to minutes, hours, etc. The main requirement - the number $N$ of consecutive trades during the averaging interval $\Delta$ (A.1) should support reliable assessments of at least the first two statistical moments of values $C(t_i)$ and volumes $U(t_i)$. The values $C(t_i)$ and volumes $U(t_i)$ of a trade at time $t_i$ determine trade price $p(t_i)$

$$C(t_i) = p(t_i)U(t_i) \tag{A.2}$$

We underline that the additive random variables of trade values $C(t_i)$ and volumes $U(t_i)$ determine statistical properties of the trade price $p(t_i)$ during the averaging interval $\Delta$ (A.1). In particular, statistical moments of random price $p(t_i)$ and price-volume correlations are determined by statistical moments and correlations of random trade values $C(t_i)$ and volumes $U(t_i)$. Moreover, it is important to highlight that probability theory (Shiryaev, 1999; Shreve, 2004) has no solution to the problem: How do given probabilities of random values $C(t_i)$ and volumes $U(t_i)$ determine the probability of random price $p(t_i)$ that obeys the equation (A.2)? Probability theory has the solution for an additive equation (A.3), but not for (A.2)

$$ln\, C(t_i) = ln\, p(t_i) + ln\, U(t_i) \tag{A.3}$$

The derivation of a probability density function (PDF) of a random variable $lnp(t_i)$ by the given PDF of random variables $lnC(t_i)$ and $lnU(t_i)$ is described in introductory notes on probability (Papoulis and Pillai, 2002; p. 181). However, the PDF of $lnp(t_i)$ doesn't help to derive the PDF of the random price $p(t_i)$ determined by the equation (A.2). We recall these known difficulties to highlight that the description of the PDF of a random price $p(t_i)$ that obeys the equation (A.2) and depends on the PDF of random trade values $C(t_i)$ and volumes $U(t_i)$ is an unsolved open problem. According to Segnitz (1870), the researchers of "statistical bureaus of several German states," and Walsh (1901; 1921), one may introduce VWAP $p(t)$ (3.1; A.6). To do that, let us define the total value $C_\Sigma(t)$ and volume $U_\Sigma(t)$ (A.4) of all trades made during the averaging interval $\Delta$ (A.1):

$$C_\Sigma(t) = \sum_{i=1}^{N} C(t_i) \quad ; \quad U_\Sigma(t) = \sum_{i=1}^{N} U(t_i) \tag{A.4}$$

Due to (A.1), there were $N$ trades made during the averaging interval $\Delta$, and we define the mean value $C(t)$ and mean volume $U(t)$ (A.5) of trades made during $\Delta$ (A.1):

$$C(t) = E[C(t_i)] = \frac{1}{N} C_\Sigma(t) = \frac{1}{N}\sum_{i=1}^{N} C(t_i) \quad ; \quad U(t) = E[U(t_i)] = \frac{1}{N} U_\Sigma(t) = \frac{1}{N}\sum_{i=1}^{N} U(t_i) \tag{A.5}$$

We follow Walsh and define "*… the average price by dividing the total sum (or the full value) by the total quantities*." (Walsh, 1921, p. 88). Decades later this definition of VWAP was reproduced by Berkowitz et al. (1988). The equations (A.2) and (A.4; A.5) determine VWAP $p(t)$:

$$p(t) = \frac{C_\Sigma(t)}{U_\Sigma(t)} = \frac{C(t)}{U(t)} = \frac{1}{\sum_{i=1}^{N} U(t_i)} \sum_{i=1}^{N} p(t_i)U(t_i) \tag{A.6}$$

We spend so much effort on the derivation of the well-known VWAP $p(t)$ (A.6) to outline a rather important statement: the average of a random price $p(t_i)$ that obeys the equation (A.2) is determined by the time series of random values $C(t_i)$ and volumes $U(t_i)$, but not by a single time series of random prices $p(t_i)$.

However, one may consider the simplified case when the mean price is determined by the price time series $p(t_i)$. Let us assume that all volumes $U(t_i)$ of consecutive trades during the interval $\Delta$ (A.1) are constant and $U(t_i)=U$. Then, from (A.6), obtain the mean price $\pi(t)$ (A.7):

$$if\ U(t_i) = U\ \ then\ \ p(t) = \frac{1}{\sum_{i=1}^{N} U(t_i)} \sum_{i=1}^{N} p(t_i)U(t_i) = \frac{1}{N \cdot U} \sum_{i=1}^{N} p(t_i)U = \frac{1}{N} \sum_{i=1}^{N} p(t_i) = \pi(t) \quad \text{(A.7)}$$

We highlight that the mean price $\pi(t)$ (A.7) has the economic meaning if and only if all volumes $U(t_i)$ of consecutive trades made during the averaging interval $\Delta$ (A.1) are constant. In this limited case the PDF of random trade value $C(t_i)$ and price $p(t_i)$ that obey the equation (A.2) are almost identical, and the PDF of a random price $p(t_i)$ is determined by the price time series only.

We repeat these trivial relations to underline a rather important statement: if one assesses the mean, variance, or PDF of a random price using the time series of prices only, then one implicitly uses the limited approximation of constant trade volumes. One should care about the consequences of applications of such a simplification of market trades. It is evident that the same statement is valid for the assessments of means, variances, or PDF of random returns.

All in all, any economic processes are the result of market trades made in the economy during the averaging interval $\Delta$. One may describe economic processes under two basic approximations. The general model of economic processes is the result of the random trades that are described by time series of random values $C(t_i)$, volumes $U(t_i)$, and prices $p(t_i)$ that obey the equation (A.2). The limited case of economic processes may be described under the simplified assumption that all volumes $U(t_i)$ of the consecutive trades made during $\Delta$ (A.1) are constant.

The economically justified assessment of price-volume correlation is the consequence of this rule. The definition of the price-volume correlation $corr[p,U]$ takes the form:

$$corr[p, U] = E[(p(t_i) - E[p(t_i)]) \cdot (U(t_i) - E[U(t_i)])] = E[p(t_i) \cdot U(t_i)] - E[p(t_i)]E[U(t_i)] \quad \text{(A.8)}$$

$E[...]$ denotes mathematical expectation. If one considers random time series of values $U(t_i)$ and prices $p(t_i)$ of trades made during $\Delta$, one should use VWAP $p(t)$ (A.6) as an average price. Mathematical expectation $E[p(t_i)]$ of price $p(t_i)$ during $\Delta$ (A.1) is determined by (A.6) and mathematical expectation $E[U(t_i)]$ of trade volume $U(t_i)$ is determined by (A.5):

$$E[p(t_i)] = p(t) = \frac{C(t)}{U(t)} \quad ; \quad E[U(t_i)] = U(t) = \frac{1}{N} \sum_{i=1}^{N} U(t_i) \quad \text{(A.9)}$$

From (A.8; A.9), (A.6), and equation (A.2), obtain:

$$E[p(t_i) \cdot U(t_i)] = E[C(t_i)] = E[p(t_i)]E[U(t_i)] \quad \text{(A.10)}$$

The use of (A.8-A.10), gives that price-volume correlation *corr[p,U]* is always zero:

$$corr[p,U] = E[p(t_i) \cdot U(t_i)] - E[p(t_i)]E[U(t_i)] \equiv 0 \quad \text{(A.11)}$$

It is evident that in the simplified case when all volumes $U(t_i)$ of the consecutive trades are constant, the price-volume correlation is all the more so identical to zero, *corr[p,U] =0*.

## Appendix B. The roots of flaws of "empirical evidence" of positive or negative price-volume correlations

The definition of correlation *corr[x,y]* between two random variables *x* and *y* has the form:

$$corr[x,y] = E[(x - E[x]) \cdot (y - E[y])] = E[xy] - E[x]E[y] \quad \text{(B.1)}$$

Almost all manuals on statistical analysis of time series (Tsay, 2005, pp. 17; 25; Montgomery et al., 2015, pp. 38, 437) give the conventional expression for the empirical calculation of correlation $corr[x,y]_e$ (B.2) between two time series $\{x_i\}$ and $\{y_i\}$ of random variables *x* and *y* in the form:

$$corr[x,y]_e = \frac{1}{N}\sum_{i=1}^{N}(x_i - E[x]) \cdot (y_i - E[y]) \quad \text{(B.2)}$$

*E[x]* and *E[y]* denote the mathematical expectations of random variables. The relations (B.2) are so common and widespread that "empirical researchers" used them without any caution with regard to specific statistical properties of a random trade price $p(t_i)$. Below we explain why the use of the conventional expression of correlation $corr[x,y]_e$ (B.2) leads to miscalculation for the case of price-volume correlation. Indeed, the use of (B.2) implies:

$$corr[x,y]_e = \frac{1}{N}\sum_{i=1}^{N} x_i y_i - e(x)E[y] - E[x]e(y) + E[x]E[y] \quad \text{(B.3)}$$

To highlight the latent problem, we define the means *e(x)* and *e(y)*:

$$e(x) = \frac{1}{N}\sum_{i=1}^{N} x_i \quad ; \quad e(y) = \frac{1}{N}\sum_{i=1}^{N} y_i \quad \text{(B.4)}$$

If the mathematical expectations *E[x]=e(x)* and *E[y]=e(y)* obey (B.4), then the expression $corr[x,y]_e$ (B.2) correctly describes the definition of correlation on the right side of (B.1):

$$corr[x,y]_e = \frac{1}{N}\sum_{i=1}^{N} x_i y_i - E(x)E[y] - E[x]E(y) + E[x]E[y] = \frac{1}{N}\sum_{i=1}^{N} x_i y_i - E(x)E[y] \quad \text{(B.5)}$$

However, if at least $E[x] \neq e(x)$, the calculation via expression (B.2) doesn't fit the definition of correlation. Let us take $p(t_i)=x_i$, $U(t_i)=y_i$, and use (B.2) and (A.5; A.6) to assess the expression $corr[p,U]_e$ (B.2) using time series of prices $p(t_i)$ and volumes $U(t_i)$:

$$corr[p,U]_e = \frac{1}{N}\sum_{i=1}^{N}(p(t_i) - p(t)) \cdot (U(t_i) - U(t)) = \frac{1}{N}\sum_{i=1}^{N} p(t_i)U(t_i) - U(t)\frac{1}{N}\sum_{i=1}^{N} p(t_i) - p(t)U(t) + p(t)U(t)$$

From the above, obtain the expression of *corr[p,U]ₑ*:

$$corr[p,U]_e = \frac{1}{N}\sum_{i=1}^{N} p(t_i)U(t_i) - \pi(t)U(t) \quad \text{(B.6)}$$

The relations (B.6) prove that the use of (B.2) doesn't conform to the definition of price-volume correlation (A.8; B.1). The origin of this violation is the presence of the mean price $\pi(t)$ (A.7) in the right side of (B.6) instead of VWAP $p(t)$ (A.6). Ultimately, from (A.5; A.6; B.6), obtain:

$$C(t) = \frac{1}{N}\sum_{i=1}^{N} C(t_i) = \frac{1}{N}\sum_{i=1}^{N} p(t_i)U(t_i) = \frac{1}{\sum_{i=1}^{N} U(t_i)}\sum_{i=1}^{N} p(t_i)U(t_i) \cdot \frac{1}{N}\sum_{i=1}^{N} U(t_i) = p(t)U(t) \quad \text{(B.7)}$$

Hence, from (B.6; B.7), finally obtain:

$$corr[p,U]_e = [p(t) - \pi(t)]U(t) \quad \text{(B.8)}$$

The expression $corr[p,U]_e$ (B.8) highlights that the use of (B.2) doesn't describe price-volume correlation but the difference between VWAP $p(t)$ (A.6) and mean price $\pi(t)$ (A.7).

The correct expression of empirical price-volume correlation *corr[p,U]* that matches the definition of correlation (A.8; B.1) has the form (B.9) and, due to (A.9-A.11) or (B.7), always equals zero:

$$corr[p,U] = \frac{1}{N}\sum_{i=1}^{N} p(t_i)U(t_i) - p(t)U(t) \equiv 0 \quad \text{(B.9)}$$

## Appendix C. The Statistical Interdependence of Prices and Volumes

Zero price-volume correlations don't mean the lack of statistical interdependence of random prices and volumes. To assess it, one may estimate the correlations $corr[p,U^2]$ between prices and squares of trade volumes or the correlation $corr[p^2,U^2]$ between squares of prices and volumes.

### C.1 The correlation $corr[p,U^2]$.

Let us denote coefficients of variation coefficients $\psi(C)$ and $\chi(U)$ (C.1) of random trade values $C(t_i)$ and volumes $U(t_i)$ and their correlation $\varphi(C,U)$ (C.1) normalized to their mean values $C(t)$ and volumes $U(t)$ (A.5) (Olkhov, 2022; 2023):

$$\psi^2(C) = \frac{cov[C,C]}{C^2(t)} \quad ; \quad \chi^2(U) = \frac{cov[U,U]}{U^2(t)} \quad ; \quad \varphi(C,U) = \frac{corr[C,U]}{C(t)U(t)} \quad \text{(C.1)}$$

$$cov[C,C] = E[(C(t_i) - C(t))^2] = E[C^2(t_i)] - C^2(t) \quad \text{(C.2)}$$

$$cov[U,U] = E[(U(t_i) - U(t))^2] = E[U^2(t_i)] - U^2(t) \quad \text{(C.3)}$$

$$corr[C,U] = E[(C(t_i) - C(t)) \cdot (U(t_i) - U(t))] = E[C(t_i) \cdot U(t_i)] - C(t)U(t) \quad \text{(C.4)}$$

$$E[C^2(t_i)] = \frac{1}{N}\sum_{i=1}^{N} C^2(t_i) \quad ; \quad E[U^2(t_i)] = \frac{1}{N}\sum_{i=1}^{N} U^2(t_i) \tag{C.5}$$

$$E[C(t_i)\cdot U(t_i)] = \frac{1}{N}\sum_{i=1}^{N} C(t_i)U(t_i) \tag{C.6}$$

Then, from (C.1-C.6), one may derive:

$$E[C^2(t_i)] = C^2(t)\left[1+\psi^2(C)\right] \quad ; \quad E[U^2(t_i)] = U^2(t)[1+\chi^2(U)] \tag{C.7}$$

Then, the use of (A.2; A.5; A.6) and (C.7) expresses *corr[p,U²]* as:

$$corr[p,U^2] = E[C(t_i)\cdot U(t_i)] - p(t)U^2(t)[1+\chi^2(U)] = C(t)U(t)[\varphi(C,U)+1] - p(t)U^2(t)[1+\chi^2(U)]$$

$$C(t)U(t) = p(t)U^2(t)$$

The above relations can be presented as:

$$corr[p,U^2] = [\varphi(C,U) - \chi^2(U)]p(t)U^2(t) \tag{C.8}$$

The relations (C.1-C.6) give the empirical expression of *corr[p,U²]* (C.8).

**C.2 The correlation *corr[p²,U²]*.**

$$corr[p^2,U^2] = E[p^2(t_i)U^2(t_i)] - E[p^2(t_i)]E[U^2(t_i)] = E[C^2(t_i)] - E[p^2(t_i)]E[U^2(t_i)] \tag{C.9}$$

The use of (C.7), gives:

$$corr[p^2,U^2] = C^2(t)[1+\psi^2(U)] - E[p^2(t_i)]U^2(t)[1+\chi^2(U)] \tag{C.10}$$

Let us introduce the coefficient of variation $\mu(p)$ (C.11) of price

$$\mu^2(p) = \frac{cov[p,p]}{p^2(t)} \quad ; \quad cov[p,p] = E\left[\left(p(t_i)-p(t)\right)^2\right] = E[p^2(t_i)] - p^2(t) \tag{C.11}$$

$$E[p^2(t_i)] = p^2(t)[1+\mu^2(t)] \tag{C.12}$$

Hence, one may present *corr[p²,U²]* (C.9) as:

$$corr[p^2,U^2] = C^2(t)[1+\psi^2(U)] - [1+\mu^2(p)]p^2(t)U^2(t)[1+\chi^2(U)]$$

$$corr[p^2,U^2] = \left[1+\psi^2(U) - [1+\mu^2(p)]\left(1+\chi^2(U)\right)\right]p^2(t)U^2(t) \tag{C.13}$$

The expression of market-based variance $\Phi(p)$ (C.14) of price has the form (Olkhov, 2022; 2023):

$$\Phi(p) = cov[p,p] = \frac{\psi^2(C) - 2\,\varphi(C,U) + \chi^2(U)}{1+\chi^2(U)}\cdot p^2(t;1) \tag{C.14}$$

Hence, the coefficient of variation $\mu(p)$ (C.11) of price takes the form

$$\mu^2(p) = \frac{\psi^2(C) - 2\,\varphi(C,U) + \chi^2(U)}{1+\chi^2(U)} \quad ; \quad 1+\mu^2(p) = \frac{1+\psi^2(C) - 2\,\varphi(C,U) + 2\chi^2(U)}{1+\chi^2(U)} \tag{C.15}$$

$$1+\psi^2(U) - [1+\mu^2(t)]\left(1+\chi^2(U)\right) = 1+\psi^2(U) - 1 - \psi^2(C) + 2\,\varphi(C,U) - 2\chi^2(U)$$

Finally, obtain:

$$corr[p^2,U^2] = 2[\,\varphi(C,U) - \chi^2(U)]p^2(t)U^2(t) = 2corr[p,U^2]\,p(t) \tag{C.16}$$

The relations (A.5; A.6; C.1; C.5; C.6) give the expressions for the empirical calculations of the correlation $corr[p,U^2]$ (C.8) and $corr[p^2,U^2]$ (C.16) using the empirical time series of the values $C(t_i)$, volumes $U(t_i)$, and prices $p(t_i)$ of the consecutive trades during $\Delta$ (A.1).